\documentclass[aps,prb,twocolumn,groupedaddress,showpacs]{revtex4}

\usepackage[dvips]{graphicx}
\usepackage[usenames,dvipsnames]{color}

\begin{document}


\title{Tunneling Characteristics of an Electron-Hole Trilayer\\
under an In-plane Magnetic Field}


\author{Y. Lin}
\author{E. E. Mendez}
\email[Corresponding author: ]{Emilio.Mendez@sunysb.edu}
\author{A. G. Abanov}
\affiliation{Department of Physics and Astronomy, SUNY at Stony Brook, Stony Brook, NY 11794-3800}

\date{\today}

\begin{abstract}
We have studied the tunneling properties of GaSb/AlSb/InAs/AlSb/GaSb heterostructures, in which electrons and holes accumulate in the InAs and GaSb regions respectively, under a magnetic field parallel to the interfaces. The low-temperature (T = 4.2K), zero-bias magnetoconductance has shown a behavior with field that evidences the two-dimensional character of both electrons and holes and that has allowed us to determine the hole density and the electron-hole separation. The observed field dependence of the current-voltage characteristics is explained by the relative change in parallel momentum of electrons and holes induced by the field. 
\end{abstract}
\maketitle
Many interesting transport phenomena have been reported on bi-layer systems formed by two-dimensional (2D) electron gases separated by a potential barrier~\cite{Mendez89,Smoliner89}, among others, Coulomb drag~\cite{Lilly98,Rojo99,Lok01} and fractional Hall states.~\cite{Eisenstein97} Lately, the interest has been on electron tunneling between the two 2D gases at high magnetic fields, at which a dramatic dc-Josephson-like behavior has been observed at very low temperatures.~\cite{Spielman01} These bi-layer systems have been implemented mostly using GaAs/AlGaAs double quantum wells, in which it is possible to make separate contact to the 2D gases and to control the carrier density of each of them.~\cite{Eisenstein90} 

The experimental results have in turn stimulated theoretical studies ~\cite{Girvin97,MacDonald01,Demler01,Mitra01}, which have even been extended to a system formed by separate 2D electron and 2D hole gases (2DEG and 2DHG, respectively) interacting with each other.~\cite{Vignale96,Conti98,Hu00} Such a system is appealing, among other reasons, because it offers the possibility of forming a gas of excitons aligned along the two parallel planes, a configuration which, according to theoretical predictions, may lead to Bose-Einstein condensation-like phenomena.~\cite{ Lozovik76a,Shevchenko76} 

Unfortunately, in the type of heterostructures exemplified by GaAs-GaAlAs it is not easy to realize a 2DEG-2DHG coupled system under equilibrium conditions. Because of the relative band alignment of these two materials, the separation between high-mobility 2DEG and 2DHG is in practice too large for exciton formation. In addition, it is difficult to make separate ohmic contacts to the two gases.~\cite{Sivan92,Kane94,Shapira99} 

An alternative materials system for the observation of 2DEG-2DHG phenomena is that formed by InAs/AlSb/GaSb. Since the GaSb valence band (top) is higher in energy than the InAs conduction band (bottom) when these two materials are brought together, negative charge is transferred from GaSb into InAs, creating holes and electrons, respectively, at the interfaces of those layers. AlSb is used as a barrier, whose thickness controls the strength of the electron-hole interaction and the amount of charge transferred between GaSb and InAs.

For vertical transport, a typical configuration is GaSb-AlSb-InAs-AlSb-GaSb, whose tunnel current-voltage (I-V) characteristic exhibits a sharp negative differential negative conductance, even at room temperature. This behavior results from the tunneling of electrons accumulated in the InAs well to hole states in the collecting GaSb electrode, followed by tunneling of valence-band electrons from the emitter into the InAs well. (Alternatively, the process can be understood in terms of holes tunneling from one electrode to the other via conduction-band states in InAs.)

The 2D character of electrons in InAs has been demonstrated by Shubnikov -- de Haas (SdH) oscillations that appear in the zero-bias magnetotunneling conductance with a field perpendicular to the heterostructure's interfaces.~\cite{Mendez91,Mendez92a} The holes, although essential to explain the tunnel I-V characteristics and their dependence on hydrostatic pressure.~\cite{Mendez92b}, are much more elusive. No direct signature of holes or their number, let alone their 2D character, has been found so far in magnetotunneling measurements.

In this work we show that the presence of 2D holes is apparent when a magnetic field is applied parallel to the interfaces (in-plane field). We have found that the zero-bias tunneling magnetoconductance exhibits two sharp maxima at fields that are unambiguously related to the presence of 2D electrons and 2D holes with different densities. The behavior of the dependence of the I-V characteristics on magnetic field is consistent with the carrier densities determined from the zero-bias measurements.

The GaSb-AlSb-InAs-AlSb-GaSb heterostructure used in this study consisted of a 150 {\AA} InAs well and 40 {\AA} AlSb barriers, with p-type GaSb electrodes. (Details of the band structure and materials preparation are found elsewhere.~\cite{Mendez91}) Previous tunneling measurements on this structure under a perpendicular field had yielded a 2D electron density of $1.2 \times 10^{12}$ cm$^{-2}$,~\cite{Mendez91} a number significantly larger than that expected from the intrinsic transfer of carriers from GaSb.~\cite{notes} 

\begin{figure}[!h]
\includegraphics[width=3in,clip=]{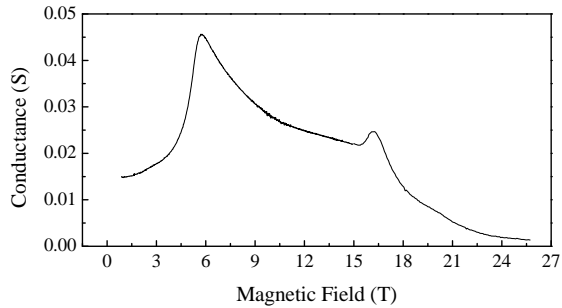}
\caption{\label{592KD}Zero-bias tunneling conductance vs in-plane magnetic field in a GaSb-AlSb-InAs-AlSb-GaSb heterostructure at T = 4.2K. The conductance shows an absolute maximum for a field of 5.8T and a local maximum at 16.3 T.}
\end{figure}

For the work described here, the tunneling current between the two p-type electrodes was measured under an in-plane magnetic field, $H$, of up to 26T and at a temperature of 4.2 K. The dependence of the zero-bias conductance on increasing field, shown in Fig.~\ref{592KD}, can be summarized as a very rapid increase at around 5 T peaking at 5.8 T, followed by a gradual decline except for a secondary peak at 16.3 T.

Similar ``double-peak'' behavior has been observed before in the tunneling magnetoconductance ofGaAs/AlGaAs double quantum wells,~\cite{Eisenstein91,Simmons93} and has been explained in terms of field-induced singularities in the tunneling conductance of two 2D electron gases. In the GaSb-AlSb-InAs-AlSb-GaSb case considered here, the situation is more complex. Thus, not only are there three tunneling regions, two of them with holes instead of electrons, but also the dimensionality of the holes is in principle unclear. 

Before analyzing the result plotted in Fig.~\ref{592KD}, the question to be addressed is then, should the magnetoconductance show any singularity when tunneling happens between a 2D and a 3D system? To answer it, it is useful to trace the origin of the conductance peaks in 2D-2D tunneling in GaAs/GaAlAs bilayers. In this case, when an in-plane magnetic field, $H_{\parallel}$, is applied the two (initially concentric) Fermi circles are shifted relative to each other by an amount $k_{H}=e\, d\, H_{\parallel}/\hbar$, where $d$ is the center-to-center distance between the two gases. The magnetoconductance can be expressed mathematically as~\cite{Lyo98}
\begin{equation}
G_{\parallel}={{4\pi e^{2}}\over {\hbar^{2}}} J_{0}^{2} \sum_{\textbf{k}}\int_{-\infty}^{\infty}[-f'(\zeta )]\rho_{1{\bf k}}(\zeta )\rho_{2{\textbf k}}(\zeta ), 
\label{G_H}
\end{equation}
where $J_{0}$ is the zero-field tunneling integral, $f(\eta)$ is the Fermi function, and $\rho_{ik}(\eta)$ is the density of states in the $i_{th}$ layer. If the case of two 2DEGs with the same density, in he limit of zero temperature and zero broadening at the Fermi level, the conductance can be simplified to
\begin{equation}
G_{\parallel}\propto \sum_{k_{x}, k_{y}} \delta(k_{F}^2-(k_{x}^2 + k_{y}^2))\ \delta(k_{F}^2-((k_{x}-k_{H})^2+k_{y}^2)),
\end{equation}
where $k_{F}$ is the Fermi wave vector. At zero field, {\textit i.e.}, $k_{H}=0$, the two delta functions are identical and the allowed values of $k_{x}$ and $k_{y}$ are innumerable; in the absence of localization, the conductance diverges. When $k_{H}\neq 0$, the allowed $k$ values are limited to those corresponding to the two intersections of the Fermi circles, and the conductance decreases. When $k_{H}=2k_{F}$ the two intersections coincide and a new divergence appears. The conductance diverges as $H_{\parallel}^{-1}$ for $k_{H}=0$ and as $H_{\parallel}^{-1/2}$ for $k_{H}=2k_{F}$. When broadening is included, these two singularities become peaks, more or less sharp depending on the broadening parameter.

If the 2DEG have different densities, and therefore different Fermi wave vectors ($\Delta k_{F}$ being the difference between them), one Fermi circle lies initially inside the other. Then the first divergence appears when $k_{H}=\Delta k_{F}$, and it is of order $H_{\parallel}^{-1/2}$ instead of $H_{\parallel}^{-1}$. Physically, the first peak observed in the magnetoconductance occurs when the smaller circle first touches the larger one; the second peak corresponds to the field at which they touch for the last time.

For 3D-2D tunneling, the summation in Eq.~\ref{G_H} ranges over the whole $\textbf{k}$ space and one of the delta functions in Eq. 2 includes the variable $k_{z}$. Because of the additional integration, if the Fermi wave vectors for the 2D and 3D gases are the same then the order of the divergence is reduced to $H_{\parallel}^{-1/2}$ when $k_{H}\rightarrow 0$ while at $k_{H}=2k_{F}$ the divergence disappears. If the two Fermi wave vectors are not the same, both divergences disappear altogether. In short, for 3D-2D tunneling the magnetoconductance should exhibit at best only one peak.

To test this prediction, we have measured the zero-bias magnetoconductance of a GaAs/GaAlAs double-barrier resonant- tunneling diode with a 2DEG in the GaAs central well. Each of the two sequential tunneling steps between the well and an electrode can be regarded as 3D-2D tunneling. As anticipated, we observed a gradual decrease of the conductance with increasing magnetic field but no peaks whatsoever.

Once we have established what to expect in the case of 2D-3D tunneling we can go back to the analysis of the magnetoconductance for GaSb-AlSb-InAs-AlSb-GaSb. The presence of two clear peaks in Fig.~\ref{592KD} is then an unequivocal indication that there are 2D holes in the GaSb electrodes. At the same time, the values of the magnetic field at which the two peaks are observed allow us to determine the ratio of 2DEG to 2DHG densities, $N_{h}/N_{e}$, as well as the effective separation, $d$, between these gases.

\begin{figure}[!h]
\includegraphics[width=3in,clip=]{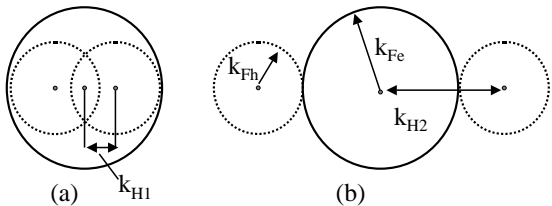}
\caption{\label{3circles}Fermi circles for 2D electrons (solid lines) and 2D holes (dotted lines) accumulated in the InAs central well and in the two GaSb electrodes, respectively, in the presence of a magnetic field parallel to the heterostructure's interfaces. (a) Relative position of the three Fermi circles at a field corresponding to the first conductance peak in Fig. 1; (b) The same as in (a) but at the field of the second conductance peak.}
\end{figure}

We start by considering three Fermi circles, two identical ones for the holes and one for the 2D electrons in the InAs well, with the radii of the former smaller than that of the latter. (In the ideal case, in which all the electrons are exclusively a result of charge transfer from GaSb, $N_{e}=2N_{h}$, but, in general, this may not be the case.) Using as a reference the electron Fermi circle, with increasing in-plane magnetic field the hole Fermi circles are shifted away in opposite directions, each by an amount, $k_{H}=e\, d\, H_{\parallel}/\hbar$, where $d$ is now the center-to-center distance between the 2D electron and hole gases (see Fig.~\ref{3circles}).

In a structure like the one discussed here, in which its symmetry makes it reasonable to assume that both hole gases have the same density, the situation becomes similar to that of two 2D gases. The first conduction peak appears at $H_{1}$, the field for which the first contact between the electron and hole Fermi circles occurs (Fig.~\ref{3circles}a), and the second peak at $H_{2}$, when those circles touch each other for the last time (Fig. 2b). From elementary geometry it is straightforward to get expressions for the $N_{h}/N_{e}$ ratio, as well as $d$, in terms of $H_{1}$ and $H_{2}$, 
\begin{equation}
\frac{N_{h}}{N_{e}} = \Big( {\frac{H_{2}-H_{1}}{H_{2}+H_{1}}} \Big) ^{2}
\end{equation}
and
\begin{equation}
d = { {4\pi \hbar N_{e}^{1/2}} \over {e(H_{1}+H_{2})} }.
\end{equation}

Using the values of $H_{1}=5.8\, T$ and $H_{2}=16.3\, T$ from Fig.~\ref{592KD} and $N_{e}=1.2\times 10^{12}$ cm$^{-2}$ determined previously,~\cite{Mendez91} we obtain $N_{h}=2.7\times 10^{11}$ cm$^{-2}$ and $d=164$ {\AA}. This value for $N_{h}$ is considerably less than the ideal $N_{e}/2$, which confirms the existence of extrinsic sources of electrons in InAs/AlSb/GaSb heterostructures. Since the determination of $N_{h}$ relies on two quantities, $H_{1}$ and $H_{2}$, measured directly, the value for $N_{h}$ should have only a small uncertainty. This is in contrast with a previous, more indirect determination that yielded a value of $4.5\times 10^{11}$ cm$^{-2}$ for $N_{h}$.~\cite{Mendez92b} This earlier number should be seen just as an estimate, as it was based on a linear extrapolation of the dependence of the peak current on the electron density, derived from hydrostatic-pressure measurements. 

As for the effective separation between the 2D electrons in InAs and the 2D holes at the GaSb interfaces, the 164 {\AA} value obtained here is quite reasonable, considering that the barrier thickness is 40 {\AA} and that half of the InAs well width is 75 {\AA}. The 49 {\AA} difference up to 164 {\AA} represents the average distance of the one-dimensional hole wavefunctions to the GaSb interface.

\begin{figure}[!h]
\includegraphics[width=3in,clip=]{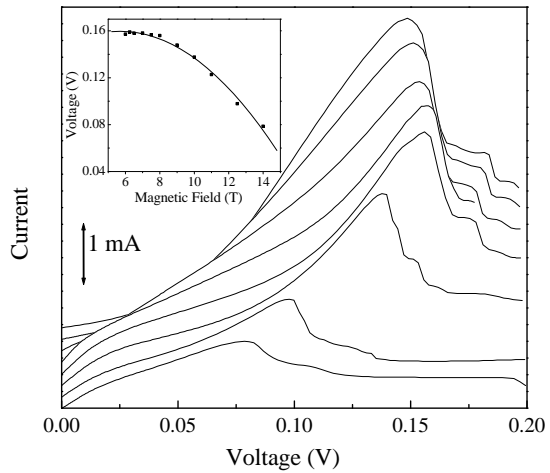}
\caption{\label{592IVH}Tunnel current-voltage characteristics for the same heterostructure as in Fig. 1, for selected in-plane magnetic fields from 0T (top curve) to 14T (bottom curve), every 2T. For clarity, the traces have been shifted vertically with respect to each other. As shown in the inset, the peak voltage changes very little with field until 6 T, and then shifts to lower values, non-linearly with the field.}
\end{figure}

The physical picture that has emerged from the zero-bias measurements -- 2D electrons and holes of different densities tunneling into each other -- is also consistent with the field dependence of the full I-V characteristics, shown in Fig.~\ref{592IVH}. With increasing field, the peak voltage varies little up to 6T, and then decreases monotonically, as summarized in the inset of Fig.~\ref{592IVH}.

\begin{figure}[!h]
\includegraphics[width=3.1in,clip=]{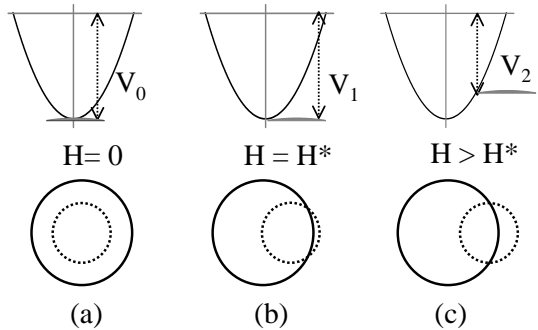}
\caption{\label{FDFCs}In-plane dispersion curve and Fermi circles for 2D electrons (solid lines) and holes (dotted lines) without a magnetic field (a), and with an in-plane field equal to (b) or larger than (c) the critical field $H^{*}$. The values $V_{0}$, $V_{1}$ and $V_{2}$, correspond, in each case, to the voltage at which the current starts to decrease because of in-plane momentum conservation. }
\end{figure}

This behavior can be understood with the help of the dispersion curves for electrons and holes, discussed previously~\cite{Mendez92a} and shown in Fig.~\ref{FDFCs} along with their respective Fermi circles. As the in-plane magnetic field increases, the hole dispersion curve "moves away" from the electron dispersion curve. Because of the much smaller hole Fermi energy, the voltage at which the current peaks, $V_{p}$, changes very little until the field reaches $H^{*}$, at which point the edge of the hole Fermi circle coincides with the center of the electron Fermi circle (see Fig.~\ref{FDFCs}b). Above that critical field, the peak voltage (which is almost identical to the voltage at which the tunnel current vanishes) decreases rapidly, following the contour of the electron dispersion curve (Fig.~\ref{FDFCs}c), and it can be expressed as
\begin{equation}
V_{H} \approx { {e\, d^{2}}\over {2m^{*}} }\, H^{2},
\label{Vb}
\end{equation}
where $m^{*}$ is the electron's effective mass.

With the values obtained above for $N_{h}$ and $d$, we can easily estimate the critical field $H^{*}$. The calculated value of 6.2T is very close to the field at which experimentally the peak voltage starts to shift (see inset of Fig.~\ref{592IVH}). Finally, by fitting the observed dependence of $V_{H}$ on $H$ (for $H\, >\, H^{*}$) to Eq.~\ref{Vb}, one can obtain $d$. Using $m^{*}=0.027m_{0}$ (which includes nonparabolicity effects) for the InAs effective mass, the fit yields $d=137${\AA}, which, although smaller, is in reasonable agreement with the 164 {\AA} value obtained above.

To summarize, from the study of the tunneling current in GaSb-AlSb-InAs-AlSb-GaSb heterostructures in a magnetic field parallel to the interfaces, we have shown that the holes that participate in the tunneling process behave as a two-dimensional gas. We have also determined, quite directly, both their density and average distance to the GaSb interfaces. The evidence of hole bi-dimensionality and the effective electron-hole separation are promising results regarding the possibility of forming equilibrium excitons with opposite charges in different planes. On the other hand, the charge imbalance we have found hinders electron-hole binding and it must be drastically reduced before one can make that promise a reality. 

We are grateful to W. I. Wang, who provided the samples used in this study. The work has been supported by the US Army Research Office.


\begin{thebibliography}{28}
\bibitem[{not()}]{notes}
\bibinfo{note}{From a self-consistent calculation of Poisson and Sch\"odinger
  equations, the intrinsic electron carrier density is estimated to be $6.8\times 10^{11}$
  cm$^{-2}$.}
\end{thebibliography}
\end{document}